\documentclass{vgtc}                          

\graphicspath{{figures/}{pictures/}{images/}{./}} 

\usepackage{times}                     

\usepackage{tabu}                      
\usepackage{booktabs}                  
\usepackage{lipsum}                    
\usepackage{mwe}                       

\usepackage{mathptmx}                  

\usepackage{enumerate}
\usepackage{amsmath,amssymb}
\usepackage{siunitx}
\onlineid{0}

\vgtccategory{Research}

\vgtcinsertpkg

\title{
    Visual Analytics of Neighborhood Attribute Profiles for Exploring Structural Equivalence
}

\author{
  Kohei Arimoto
  \thanks{
    e-mail: khyarmt@plusf.jp
  }\\ %
  \scriptsize Teikoku Databank, Ltd. %
  \and
  Masahiko Itoh
  \thanks{
    e-mail: imash@do-johodai.ac.jp
  }\\ %
  \scriptsize Hokkaido Information University %
}

\teaser{
    \centering
    \includegraphics[width=\linewidth]{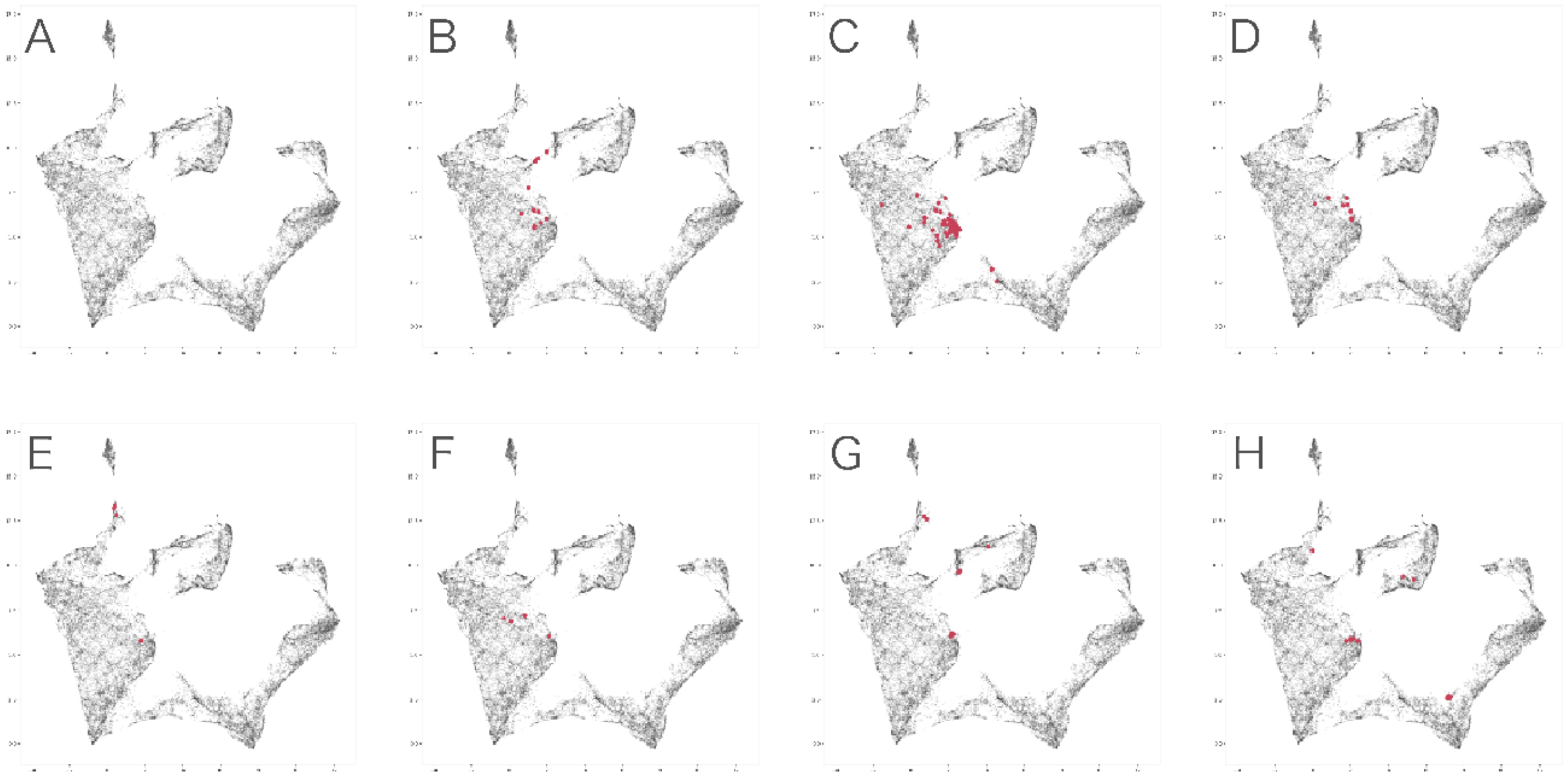}
    \caption{
         Visual verification of the high-dimensional feature space of neighborhood attribute profiles and industry classifications in an inter-firm transaction network.
          (A) UMAP projection results for all nodes. The data forms a complex, non-linear manifold accompanied by density biases.
          (B–H) Overlaid display of seven manufacturing categories (B: Steel, C: Auto Parts, D: Automobiles, E: Trucks, F: Motorcycles, G: Shipbuilding, H: Construction Machinery).
          The following three characteristics can be observed from the visualization results: 
          (1) The supply chain hierarchy (B: Steel $\rightarrow$ C: Auto Parts $\rightarrow$ D: Automobiles) transitions continuously across the manifold. 
          (2) Transportation equipment (D, E, F, G), which is often treated identically under general industry classifications, is clearly separated into different regions due to differences in actual transaction networks.
          (3) Construction Machinery (H), which shares a single industry label, is fragmented into multiple regions.
          This visual evidence challenges the intuitive assumption that identical semantic classifications imply similar structural roles, suggesting that interpreting the manifold's topology may be crucial for evaluating true similarity. 
    }
    \label{fig:scatter_plots_comparison}
}

\abstract{
  Exploring similar nodes in attributed networks represents a key challenge in data mining.
  While recent representation learning methods embed networks into low-dimensional vectors, they often implicitly assume a uniform and continuous feature space.
  This paper proposes a visual analytics approach using dimensionality reduction to help clarify the true topological structure of high-dimensional feature spaces formed by nodes' neighborhood attribute profiles.
  Analyzing inter-firm transaction networks indicates that structural roles can form complex, non-linear manifolds with density biases.
  Comparing this feature space with industry classifications suggested: (1) supply chain hierarchies transition continuously; (2) categories treated identically under general semantics can be clearly separated by actual transaction networks; and (3) a single industry label may fragment into multiple regions.
  These findings suggest potential limitations in assuming identical semantics imply similar structural roles and highlight the possible need for new similarity metrics aligned with manifold topology.
} 


\begin{document}

\firstsection{Introduction}
\maketitle

Attributed networks are frequently observed in various application domains and take diverse forms.
The data ranges from simple topological information (e.g., direct links in social graphs) to integrations of rich metadata held by each node (e.g., categorical variables like industry type or size in inter-firm transactions).
In many cases, these data are composed of combinations of complex connections and diverse attribute variables.

Based on such attributed networks, users often need to explore data and attempt to answer complex structural questions.
For instance, consider an analyst searching an inter-firm transaction network for potential customers or competitors that share characteristics similar to a specific company.
To extract true value from this type of data, it is often necessary to (1) capture the roles within the network beyond mere direct connections (proximity), (2) analyze the attribute composition of neighboring nodes (e.g., the proportion of transactions with specific industry groups), and (3) distinguish and compare the general semantic similarity of business activities against true structural equivalence based on network topology.

Against the backdrop of the importance of similar node exploration, a wide variety of methods have been proposed to represent roles on a network.
These include proximity-based approaches like SimRank\cite{jeh2002simrank}, as well as representation learning methods represented by Node2Vec\cite{grover2016node2vec} and Graph Convolutional Networks (GCN)\cite{kipf2017semi}.
However, these methods can face limitations, such as overfitting to local relationships (e.g., having common neighbor nodes) and implicitly assuming that the overall space is uniform and continuous, treating the latent space geometry of the generated high-dimensional feature space as a black box.
Furthermore, many analytical methods tend to rely heavily on assigned static attribute labels.

This paper presents a visual analytics approach that leverages dimensionality reduction techniques to help uncover the true topological structure of the high-dimensional feature space formed by the nodes' neighborhood attribute profiles.
This approach can be applied to network data with complex attribute distributions and heterogeneous structural roles.
Additionally, it aims to challenge the constraints of uniform linear spaces implicitly assumed by existing methods and the premise of homogeneity between attribute labels and structure, visually demonstrating the potential need for novel similarity evaluations along non-linear manifolds.
Specifically, this research aims to offer the following three main contributions:

\begin{itemize}
    \item Elucidation of the manifold structure of structural roles: We illustrate that the neighborhood attribute profiles of nodes may not merely form discrete clusters or uniform spaces, but rather complex non-linear manifolds with continuous gradients accompanied by density variations.  Demonstration of the divergence between semantic classification (convention) and structural role (reality): We point out that the general assumption that identical semantic classifications imply similar structural roles might not always hold true.
    \item We define the risk that relying on static attribute labels could lead to misinterpretations of the true structural equivalence of a network, presenting a fundamental issue to consider in algorithm design.  Visual demonstration of structural characteristics through application examples: In a case study using actual inter-firm transaction networks, we show that (1) supply chain hierarchies transition continuously, (2) similar industries such as transportation equipment can be clearly separated by differences in transaction networks, and (3) a single industry label can be fragmented into multiple regions.
    \item This demonstrates the potential advantages and utility of our visual analysis, which does not rely solely on attribute labels.
\end{itemize}

\section{Related Work}
Exploring similar nodes in attributed networks has been a major research topic in data mining and network analysis for many years.
Initial approaches, represented by Common Neighbors and SimRank\cite{jeh2002simrank}, were generally based on proximity, such as direct links or paths between nodes.
Subsequently, network embedding methods based on random walks, such as DeepWalk\cite{perozzi2014deepwalk} and node2vec\cite{grover2016node2vec}, emerged, enabling the mapping of network topology into low-dimensional vector spaces.
More recently, Graph Neural Networks (GNNs) like Graph Convolutional Networks (GCN)\cite{kipf2017semi} and GraphSAGE\cite{hamilton2017inductive} have been widely adopted as powerful frameworks for simultaneously learning node connection information and attributes.
However, many of these representation learning methods tend to evaluate similarity using global linear metrics, such as Euclidean distance or cosine similarity, in the generated high-dimensional feature spaces.
These calculation methods appear to rely on the implicit assumption that the latent space is geometrically uniform and continuous.

Research on extracting positions and roles within networks is also closely related to the problem setting of this study.
Methods like RolX\cite{henderson2012rolx} and struc2vec\cite{ribeiro2017struc2vec}, which identify roles such as hubs or peripheral nodes using network motifs and degree distributions without overfitting to local link sharing, have brought important progress in capturing structural equivalence\cite{lorrain1971structural}.
Furthermore, numerous approaches have been proposed to characterize structural and semantic roles by aggregating the categorical attributes of the node itself and its neighbors.
While the neighborhood attribute profiles evaluated in this study belong to this lineage, our approach is quite different in that we question the geometric validity of the feature space itself through visual analysis, whereas existing studies often apply these feature vectors as black boxes for similarity calculations.

Additionally, many existing frameworks for representation learning and similarity exploration tend to implicitly trust assigned semantic labels, such as industry classifications, as the absolute ground truth for evaluating node homogeneity.
However, visual analytics plays an essential role in helping to elucidate the topological structure of high-dimensional spaces and in opening the black boxes of machine learning models.
Non-linear dimensionality reduction techniques like t-SNE\cite{vandermaaten2008visualizing} and UMAP\cite{mcinnes2018umap} project high-dimensional data into low-dimensional spaces while aiming to preserve local structures, and have been widely utilized to understand and evaluate model behavior\cite{smilkov2016embedding, jin2022gnnvis, wang2020understanding}.
This study applies these visual approaches to help elucidate the non-linear geometric characteristics\cite{nonato2018multidimensional, espadoto2019toward} inherent in the feature space itself, as well as to visually clarify the divergence that can occur between static attribute labels and actual structural roles\cite{heimerl2018interactive}.
By demonstrating the potential fragmentation of roles hidden within the same attribute label and the continuity of supply chains spanning different labels, we raise questions that challenge algorithm design assumptions and overreliance on semantic labels, presenting a new form of contribution to the visual analytics field.

\section{Datasets}

\begin{figure}[t]
    \label{inter_firm_transaction_network}
    \includegraphics[width=\linewidth]{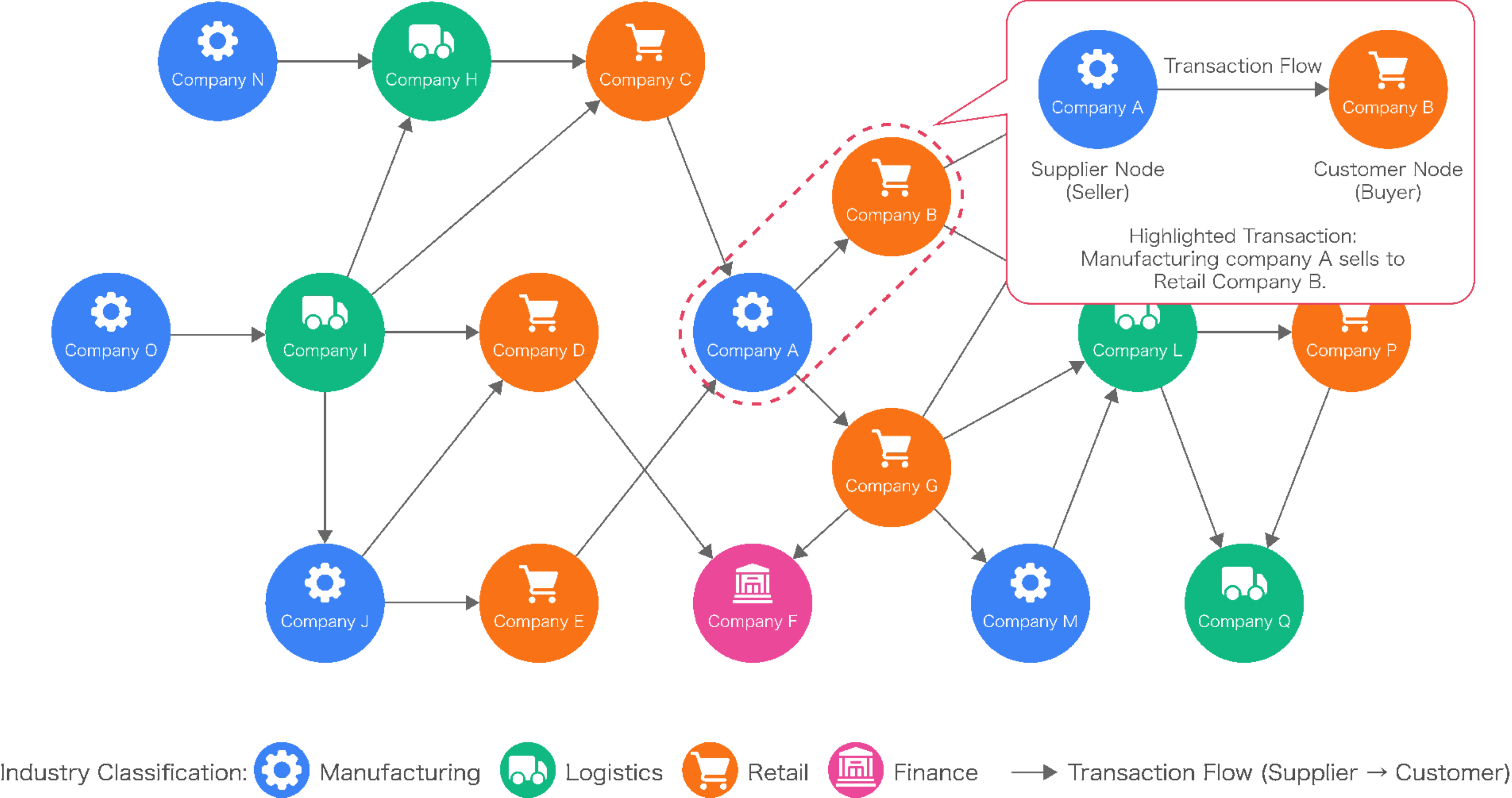}
    \caption{
        Conceptual diagram of the inter-firm transaction network.
        Nodes represent companies, with their colors and icons corresponding to industry classifications (e.g., manufacturing, logistics, retail, finance).
        Directed edges (arrows) indicate the transaction flow from suppliers to customers.
        The pop-up on the top right highlights a specific transaction relationship from Company A (Supplier) in the manufacturing industry to Company B (Customer) in the retail industry.
    }
\end{figure}

This section describes the datasets used in this study and the structural characteristics of the target inter-firm transaction network.
In particular, we aim to clarify the challenges regarding how the scale-free property of the network can introduce biases into existing similar node exploration methods.

\subsection{Inter-firm Transaction Network}
To evaluate the practicality and versatility of the proposed method, this study uses two types of large-scale datasets constructed by a major credit research agency in Japan.
These datasets cover all industries in Japan and are characterized by comprehensively including small to large enterprises without filtering by company size.
Fig. \ref{inter_firm_transaction_network} illustrates the conceptual diagram of the inter-firm transaction network handled in this study.
First, there is network structure data based on transactions between companies.
We define this network as a directed graph $G=(V, E)$.
Here, $V$ is a set of unique companies (nodes) represented by circles in the figure, and $E$ is a set of transaction relationships (edges) represented by arrows connecting the nodes.
In our dataset, this network consists of $|V|=$ \num{887289} nodes and $|E|=$ \num{6725600} directed edges.
As highlighted in the pop-up at the top right of Fig. \ref{inter_firm_transaction_network}, a directed edge $e_{ij}=(v_i, v_j) \in E$ indicates that there is a commercial flow involving the provision of products or services from company $v_i$ (Supplier Node / Seller) to company $v_j$ (Customer Node / Buyer). 

Second, there is company attribute data recording the metadata of each node $v \in V$.
In this study, we specifically focus on industry classification, which systematically categorizes the business content of companies.
As shown in the legend at the bottom of Fig. \ref{inter_firm_transaction_network}, the color and icon of each node represent the primary industry to which the company belongs (e.g., a blue gear icon is Manufacturing, an orange cart icon is Retail).
We define the set of all 13 industry categories as $C = \{c_A, c_B, ..., c_M\}$, and assume that each company node is uniquely assigned a corresponding category $c(v) \in C$. 

\subsection{Structural Features and Challenges}
\begin{figure}[t]
    \label{degree_distribution}
    \includegraphics[width=\linewidth]{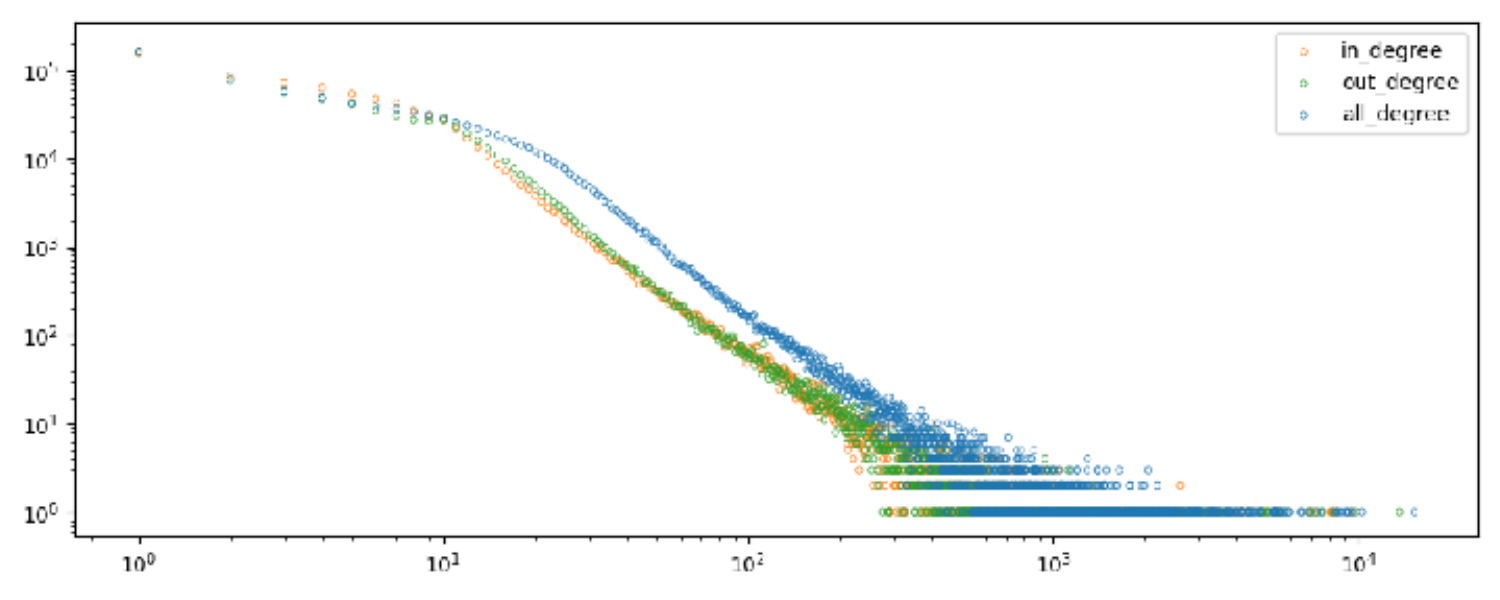}
    \caption{
        Degree distribution of the inter-firm transaction network (log-log plot).
        Blue, orange, and green indicate the total degree, in-degree, and out-degree, respectively.
        The distribution exhibits a linear decay trend, suggesting that the network possesses a scale-free property following a power law.
    }
\end{figure}

To quantitatively clarify the topological characteristics of the inter-firm transaction network, we analyzed the degree distribution.
Here, we define the in-degree (number of purchasing transactions) and out-degree (number of sales transactions) for each node as $k_{in}(v)$ and $k_{out}(v)$, respectively.
Fig. \ref{degree_distribution} plots the distributions $P(k)$ of in-degree, out-degree, and total degree across the entire dataset on a log-log graph.
The graph shows a clear linear trend, suggesting that the degree distribution follows a power law: $P(k) \propto k^{-\gamma}$.
This result indicates that the target inter-firm transaction network may possess a typical scale-free property.
In other words, a significant structural heterogeneity was observed where a very small number of hub companies have an enormous number of transaction partners, while the vast majority of companies have very few transactions.

This scale-free property introduces a major technical challenge in the similar node exploration task addressed in this study.
If conventional similarity metrics based on adjacency (such as Jaccard coefficient or cosine similarity) or basic graph embedding methods are applied directly, hub nodes with high degrees tend to dominate the calculation results.
Consequently, there is a risk of bias where hub companies might show high similarity to almost all nodes simply because of the quantitative reason of connecting with many companies, even if the qualitative similarity in business content or transaction patterns is low.

\subsection{Community Structure and Network Distance}
Furthermore, inter-firm transaction networks tend to have high modularity.
Companies are usually strongly influenced by geographical proximity or supply chain affiliations centered around specific final product manufacturers (OEMs), forming dense clusters internally.
This topological locality can cause a serious issue in similar node exploration, namely the divergence between distance on the network and functional similarity.
For example, consider two companies, A and B, that manufacture exactly the same type of auto parts.
If Company A belongs to a cluster in Region X, Affiliation Y and Company B belongs to a cluster in Region Z, Affiliation W, their business contents and roles might be very similar, yet their distance on the network topology (number of hops) will be quite large.

Existing graph embedding methods like DeepWalk and Node2Vec strongly rely on the homophily assumption, which attempts to place nodes that are close on the network close to each other in the embedding space.
Therefore, while these methods may excel at capturing similarity within the same community, they can inherently struggle to detect nodes that are structurally distant but functionally similar as roles, as in this example.
Thus, to help evaluate global similarity without relying on the local connection structure of the network, the approach of dimensionality reduction based on attribute profiles proposed in the next section can be highly relevant.

\section{Methodology}
The structural role of a node in this study can be defined and formulated as a neighborhood attribute profile.
This profile aggregates the attribute distribution and transaction direction of adjacent nodes, remaining largely independent of the distance on the network.
The inter-firm transaction network is modeled as a directed graph $G=(V, E)$ accompanied by strong degree imbalance (scale-free property).
For each company (node) $v \in V$, we define the in-degree neighborhood $N^{in}(v) = \{u | (u, v) \in E\}$ and the out-degree neighborhood $N^{out}(v) = \{u | (v, u) \in E\}$.
To ensure the statistical reliability of the neighborhood profiles, we restricted our analysis to nodes with a total degree of 50 or greater (i.e., $|N^{in}(v)| + |N^{out}(v)| \geq 50$), resulting in a final sub-network of \num{33299} nodes.
Assuming the number of industry categories is $K$, the initial feature vector $h_v \in \mathbb{R}^{2K}$ representing the transaction pattern of a node is defined as the number of connections with each industry in both input and output directions.
Here, $x_{v,k}^{in}$ represents the number of nodes belonging to industry $k$ among $N^{in}(v)$, and $x_{v,k}^{out}$ represents the number of nodes belonging to industry $k$ among $N^{out}(v)$.
This representation explicitly separates and retains the company's trading partner composition and the direction of the commercial flow.

Next, we perform preprocessing to help eliminate the bias due to company size (degree) peculiar to scale-free networks.
To prevent hub companies with large degrees from dominating the visualization results, we perform L1 normalization on $h_v$ using the total degree of node $v$, $d_v = |N^{in}(v)| + |N^{out}(v)|$, to generate a proportion vector $\tilde{h}_v = \frac{h_v}{d_v}$.
This vector $\tilde{h}_v$ expresses the pure composition ratio of transaction patterns independent of the number of connections, providing potential robustness against company size (scale-free robustness).
To help elucidate the topology of the normalized high-dimensional feature space, we apply UMAP, a dimensionality reduction method, to project it into a 2D space.
Linear dimensionality reduction like PCA depends on variance structure and can be easily dominated by outliers (high-degree nodes), while t-SNE excels at preserving local structures but can result in unstable global macrostructures.
UMAP is arguably more suitable for the purpose of this study because it can project stable global structures into lower dimensions while helping to preserve local neighborhood structures in high-dimensional spaces. 
This methodological choice is grounded in a previous comparative analysis\cite{arimoto2024dimensionality}, which evaluated multiple linear and non-linear dimensionality reduction techniques for this specific network structure and demonstrated the superior effectiveness of UMAP.

Furthermore, we applied cosine distance as the internal metric for UMAP.
By using cosine distance on the proportioned vectors, we can evaluate the similarity of the angles formed by the connection composition ratio vectors, rather than the apparent distance in Euclidean space.
The projection result of UMAP optimized by this combination of proportioning in/out-degrees and cosine distance aims to eliminate distortions caused by mere scale differences and serves as a foundation to visualize the true structural equivalence inherent in the network.

\section{Results and Discussion}

In this section, we apply UMAP to project the 26-dimensional neighborhood attribute profiles, which aggregate the attribute-specific degrees of adjacent nodes for each node, into a 2D space to evaluate their distribution structure.
To appropriately preserve the continuity of the global manifold and the contrast in local density, we set the number of neighbors ($n\_neighbors$) to 50 and the minimum distance ($min\_dist$) to 0.0.
These specific values were empirically selected based on preliminary experiments exploring various parameter combinations, aimed at maximizing both visual clarity and the distinct separation of local structural roles.

\subsection{Geometric Interpretation of Manifold Structure}
The 2D scatter plot obtained by the proposed method (Fig. \ref{fig:scatter_plots_comparison}) does not merely show a collection of nodes, but rather appears to form a non-linear manifold with complex geometric features.
Proximity in this space reflects structural equivalence, meaning the attribute composition ratios of neighboring nodes are similar, rather than the hop distance on the network.
Looking at the space as a whole, the data does not distribute uniformly; instead, a hierarchical density structure can be observed, consisting of extremely high-density core regions, low-density transition regions (bridges) that connect them, and tail regions extending from the edges of the manifold.
The high-density cores correspond to accumulations of node groups that play typical structural roles (e.g., standard transaction patterns in a supply chain) that statistically appear frequently in the target network.
On the other hand, the low-density bridge regions suggest nodes with intermediate properties of multiple roles, visually indicating that structural roles in the network might exist as continuous gradients rather than discrete clusters.

\subsection{Visual Verification of Structural Roles Based on Industry Classification}
To qualitatively evaluate how well the proposed neighborhood attribute profile and manifold space capture true structural equivalence in real-world networks, we examined the spatial distribution of nodes belonging to specific industries.
In evaluating the similar node exploration task, we constructed ground truth data for the same industry based on a comprehensive industry map.
While standard databases often assign strict single industry labels to each company based on certain rules, this comprehensive industry map attempts to capture complex and multifaceted connections, potentially enabling the evaluation of similarity discoveries.

Fig. \ref{fig:scatter_plots_comparison} shows the result of superimposing seven different manufacturing subcategories (B: Steel, C: Auto Parts, D: Automobiles, E: Trucks, F: Motorcycles, G: Shipbuilding, H: Construction Machinery) on the feature space.
From visual analysis, the following important characteristics emerged, challenging the implicit assumption that if semantic classifications are the same, structural roles should also be similar.

First, a continuous transition of the supply chain hierarchy was observed.
The node groups of Steel (B), Auto Parts (C), and Automobiles (D) are adjacent to each other in the massive high-density core region on the left side of the manifold, forming a continuous distribution.
This suggests that the actual, robust supply chain structure from materials to parts and final products can be reconstructed not just as discrete clusters, but as a continuous gradient of positions on the manifold.

Second, a clear separation between semantic classification (convention) and structural role (reality) was shown.
Automobiles (D), Trucks (E), Motorcycles (F), and Shipbuilding (G) are all generally assigned transportation equipment in industry classifications, and they can easily be treated identically as functionally similar nodes if simply filtered by industry classification.
However, on the manifold, while automobiles are located in the core region, trucks are mapped to the tail region on the lower right, and shipbuilding is clearly separated into the lower transition region.
This indicates that even if the product concepts are similar, the proposed method can help sharply separate them as completely different structural roles if the attribute distributions of actual trading partners or the nature of commercial flows differ.

Third, the possible fragmentation of roles hidden within the same industry label was highlighted.
Generally, node groups with the same industry label are assumed to behave similarly on a network, but the node group of Construction Machinery (H) does not form a single cluster; instead, it is mapped in a fragmented manner across multiple distant regions on the manifold.
This suggests the fact that even with the same static attribute label, subgroups with entirely different transaction networks may coexist in the actual network.

These visual evidences indicate the potential limitations of existing methods that rely on simple attribute label matches or linear distances in high-dimensional spaces.
This experiment supports the idea that in order to evaluate the true similarity of nodes, an interpretation aligned with the manifold's topology may be helpful, without being misled by apparent attribute labels.

\section{Conclusion}
In this paper, we proposed a visual analytics approach using neighborhood attribute profiles and UMAP to help challenge implicit assumptions in existing similar node exploration methods.
Dimensionality reduction considering scale-free graph characteristics revealed that the high-dimensional feature space may not be a simple linear space, but rather a non-linear manifold with density biases reflecting real-world constraints.
This visual evidence suggests caution regarding representation learning models that assume spatial uniformity or global linear distances.

Furthermore, our analysis demonstrated a significant potential divergence between intuitive industry classification (semantics) and actual network structure (topology).
Findings such as continuous supply chain transitions and the fragmentation of roles within the same industry label challenge the notion that identical semantics imply similar structural roles.
These outcomes suggest that evaluating true node similarity may require interpreting the manifold's geometric shape rather than relying solely on apparent attributes.
Future work includes exploring algorithms that calculate similarity geodesically along these continuous manifolds and developing interactive visual analytics systems to explore the divergence between semantics and structural roles.


\bibliographystyle{abbrv-doi}

\bibliography{bibliography}
\end{document}